\documentclass[preprint]{aastex}
\input epsf

\newcommand\bs{\hspace{-.25cm}}
\newcommand\bss{\hspace{-.07cm}}

\newcommand\tablek         {1}
\newcommand\tablem         {2}
\newcommand\tablemm        {3}
\newcommand\tablekegs      {4}
\newcommand\tablecpu       {5}
\newcommand\tableram       {6}
\newcommand\tablesim       {7}

\newcommand\figerror       {1}
\newcommand\figfftw        {2}
\newcommand\figdiv         {3}
\newcommand\figv           {4}
\newcommand\figspectra     {5}

\newcommand\B{{\bf B}}

\newcommand\kk{{\bf k}}

\newcommand\V{{\bf V}}

\newcommand\bnabla{\mbox{\boldmath $\nabla$}}

\newcommand\spa{\scriptscriptstyle\mathsf{Pade}}

\begin{document}

\nonumber
\setcounter{equation}{0}

\title{\mbox{High-Wavenumber Finite Differences and Turbulence Simulations}}

\author{Jason Maron} \affil{Department of Physics, University of
Rochester and University of Iowa}

\altaffiltext{1}{maron@tapir.caltech.edu}

\begin{abstract}

We introduce a fluid dynamics algorithm that performs with nearly spectral
accuracy, but uses finite-differences instead of FFTs to compute gradients and
thus executes 10 times faster.  The finite differencing is not based on a
high-order polynomial fit. The polynomial scheme has supurb accuracy for
low-wavenumber gradients but fails at high wavenumbers. We instead use a scheme
tuned to enhance high-wavenumber accuracy at the expense of low wavenumbers,
although the loss of low-wavenumber accuracy is negligibly slight. A tuned
gradient is capable of capturing all wavenumbers up to 80 percent of the
Nyquist limit with an error of no worse than 1 percent. The fact that gradients
are based on finite differences enables diverse geometries to be considered and
eliminates the parallel communications bottleneck.

\end{abstract}

\section{Introduction}

The spectral fluid algorithm (Canuto et. al. 1987) uses Fast Fourier Transforms
(FFTs) to compute gradients, the most precise means possible. Finite-difference
gradients based on a polynomial fit execute faster than FFTs but with less
accuracy, necessitating more grid zones to achieve the same resolution as the
spectral method. The loss of accuracy outweighs the gain in speed and the
spectral method has more resolving power than the finite-difference method.  We
introduce an alternative finite-difference formula not based on a polynomial
fit that executes as quickly but with improved accuracy, yielding greater
resolving power than the spectral method.

In section \ref{finite} we derive high-wavenumber finite difference formulas
and exhibit their effect on the resolving power of a turbulence simulation in
section \ref{flop}. In section \ref{mhd} we apply finite differences for the
purpose of mimicking the spectral algorithm, and then proceed to other
applications in section \ref{applications}.

\section{Finite differences} \label{finite}

Define a function $f_j(x_j)$ on a set of grid points $x_j = j$ with j an
integer. Then construct a gradient $f^\prime(0)$ at $x=0$ from sampling a
stencil of grid points with radius (or order) $S$ on each side. The familiar
result for the gradient on a radius-1 stencil is $f^\prime(0) \sim (f_{1} -
f_{-1})/2,$ which is obtained from fitting a polynomial of degree $2$ to $f_j.$
For a degree 4 polynomial on a radius-2 stencil, $$f^\prime(0) \sim
\frac{1}{12} f_{-2} - \frac{2}{3} f_{-1} + \frac{2}{3} f_1 - \frac{1}{12}
f_2.$$ For a stencil of order $S$, \begin{equation} f^\prime(0) \sim
\sum_{j=-S}^{S} M_j f_j \label{stencil} \end{equation} where $M_{-j} = -M_j.$

Consider the finite-difference error at x=0 for a Fourier mode $\sin(\pi k x).$
Cosine modes can be ignored because they don't contribute to the derivative at
$x=0.$ Note that the wavenumber $k$ is scaled to grid units so that $k=1$
corresponds to the maximum wavenumber expressible on the grid, also known as
the Nyquist limit. The finite difference formula (\ref{stencil}) gives
$f_k^\prime(0) \sim 2 \sum_{j=1}^m M_j \sin(j k ).$ Whereas the correct value
should be $k \pi,$ define an error function $$E_S(k) = k \pi - 2 \sum_{j=1}^S
M_j \sin(\pi k j). \label{eqerror}$$ Figure $\figerror$ shows $E_S(k)$ for
stencils of radius 1 through 24. The first order finite difference formula is
quite lame, delivering 1 percent accuracy only for $k$ less than $0.12.$
Brandenburg (2001) recognized that higher-order finite differences can
significantly extend the wavenumber resolution of the polynomial-based finite
difference.  The 8th order finite difference accuracy is better than 1 percent
up to $k=.56$ and the 16th order finite difference up to $k=.66.$ Nevertheless
these are still far from the Nyquist limit of $k=1$ and even higher-order
finite-differences yield little further progress.

A Fourier transform gives the correct gradient for all $k$ up to unity. This is
why the spectral algorithm delivers the most resolution per grid element. The
resolution limit is set by the maximum $k$ for which gradients can be captured
accurately. Although the 8th order finite-difference formula involves
considerably fewer floating point operations than an FFT, the loss of
resolution still renders it less efficient than the spectral method.

Polynomial-based finite differences have high accuracy at low $k$ but fail at
large $k.$ We can instead construct a more practical scheme to improve high-$k$
accuracy at the expense of low-$k$ accuracy, yet the loss of low-$k$ accuracy
is negligibly small. From equation \ref{eqerror}, we see that the problem of
computing accurate gradients reduces to optimizing, or ``tuning" the
coefficients $M_j$ to minimize the error over a suitable range of k, or
equivalently to construct a sine series that best mimics a linear function. A
set of tuned coefficients appear in table \tablem \hspace{1mm} with the
associated error functions in figure \figerror. The error in the radius-8 tuned
finite difference is less than 1 percent up to $k=.80,$ a dramatic improvement
over the radius-8 polynomial.  An algorithm based on tuned gradients still has
a lower maximum $k$ than the spectral algorithm but due to its increased speed
it has greater resolving power (section \ref{flop}). Henceforth we denote these
tuned gradients as ``hypergradients."

\subsection{Tuning the hypergradient coefficients}

Minimizing the error function involves a multiparameter optimization of the
coefficients $M_j,$ - a problem-dependent task with multiple legitimate
options. In fact, a high degree of customization is possible for the form of
the error function. For this application we proceed as follows. Define a target
$k_{max}$ and an indicator for the quality of the tuned coefficients: $$ E =
\int_0^{k_{max}} dk \left[ \pi k - 2 \sum_{j=1}^{m} M_j \sin( \pi j k)
\right]^4.$$ Then, perform a multi-dimensional optimization on $M_j.$ The use
of a fourth power ensures an approximately uniform error within $0 < k <
k_{max},$ although a weight function could be added to further customize the
form of the error function. $k_{max}$ is then adjusted until the error is 1
percent. The procedure is repeated for each order $S$ to yield the coefficients
in table \tablem.

It is worth noting that the radius-8 tuned coefficients are similar to the
radius-24 polynomial coefficients. This is not surprising because the
polynomial coefficients are too small to matter outside of radius 8.

\begin{center}
\begin{tabular}{|llllllllll|}
\hline
 Stencil radius&       1   & 2   & 3   & 4   & 5   & 6   & 8   & 16  & 24  \\
\hline
 Polynomial          & .12 & .24 & .34 & .40 & .44 & .48 & .56 & .66 & .72 \\
 Hypergradient       & .20 & .38 & .54 & .64 & .70 & .74 & .80 & .84 & .92 \\
\hline
\end{tabular}
\end{center}

{\it Table \tablek: Maximum resolved wavenumber $K$ for the
polynomial and hypergradient finite differences shown in figure \figerror. The
tolerance in the relative error is 1 percent.}

\begin{center}
\begin{tabular}{|lrrrrrrrrr|}
\hline
Operation &\bs $M_0$&\bs $M_1$&\bs $M_2$&\bs $M_3$&\bs $M_4$&\bs $M_5$&
\bs $M_6$&$M_7$&$M_8$\\
\hline

$\partial/\partial x$ (T1)& \bs 0 &\bs .5245&\bs &\bs &\bs 
\bs &\bs &\bs &\bs &\bs \\

$\partial/\partial x$ (T2)& \bs 0 &\bs .73694&\bs -.12780&\bs &\bs 
\bs &\bs &\bs &\bs &\bs \\

$\partial/\partial x$ (T3)& \bs 0 &\bs .83793&\bs -.23758&\bs .05000&
\bs        &\bs &\bs &\bs &\bs \\

$\partial/\partial x$ (T4)& \bs 0 &\bs .89562&\bs -.31797&\bs .11245&
\bs -.02850 &\bs &\bs &\bs &\bs \\

$\partial/\partial x$ (T5)& \bs 0 &\bs .92767&\bs -.36886&\bs .16337&
\bs -.06469 &\bs .01872&\bs &\bs &\bs \\

$\partial/\partial x$ (T6)& \bs 0 &\bs .94453&\bs -.39729&\bs .19577&
\bs -.09322 &\bs .03843&\bs -.01194&\bs &\bs \\

$\partial/\partial x$ (T8)&0&\bs .96890&\bs -.44249&\bs .25170&\bs -.15066&
\bs .08825&\bs -.04856&\bs .02376&\bs -.01051\\


$\partial^2/\partial x^2$ \bss (T8)&\bs -3.25820&\bs 1.97177&\bs -.47293&
\bs .19568& \bs -.10008&\bs .05565&\bs -.03201&\bs .01821&\bs -.01209 \\

$\partial^4/\partial x^4$ \bss (T6)&\bs 16.36332&\bs -12.46631&\bs 5.65080&
\bs -1.78786&\bs .53027&\bs -.12637&\bs .01729&\bs &\bs \\

Interp. (T8)&\bs 0 &\bs .63108&\bs -.20002&\bs .10804&\bs -.06614& \bs
.04104&\bs -.02501&\bs .01406&\bs -.00815\\

$\partial/\partial x$\,(P1)&\bs 0&\bs .50000&\bs &\bs &
\bs &\bs &\bs &\bs &\bs \\

$\partial/\partial x$\,(P2)&\bs 0&\bs .66667&\bs -.08333 & \bs &
\bs &\bs &\bs &\bs &\bs \\

$\partial/\partial x$\,(P3)&\bs 0&\bs .75000&\bs -.15000 &\bs .01667&
\bs &\bs &\bs &\bs &\bs \\

$\partial/\partial x$\,(P4)&\bs 0&\bs .80000&\bs -.20000 &\bs .03810&
\bs-.00357 &\bs &\bs &\bs &\bs \\


$\partial/\partial x$\,(P8)&\bs 0&\bs .88889&\bs -.31111&\bs .11313&
\bs -.03535&\bs .00870&\bs -.00155&\bs .00018&\bs -.00001\\



\hline
\end{tabular}
\end{center}

{\it Table \tablem: The finite-difference coefficients $M_j$. The label
``$P(S)$" corresponds to the polynomial-based finite difference on a radius $S$
stencil. ``$T(S)$" denotes the tuned hypergradient finite difference
coefficients engineered to have a relative error of 1 percent. The entry
labeled ``Interp" is for the tuned coefficients used in interpolating halfway
between two grid points. The entry labeled ``Pad\'e" gives the coefficients for
the Pad\'e derivative in explicit finite-difference form.}

\begin{center}
\begin{tabular}{|lrrrrrrrrr|}
\hline
Operation &\bs $M_0$&\bs $M_1$&\bs $M_2$&\bs $M_3$&\bs $M_4$&\bs $M_5$&
\bs $M_6$&$M_7$&$M_8$\\
\hline

$\partial/\partial x$ (T16)&\bs 0&\bs .98043&\bs -.46188&\bs .27843&\bs -.18085
&\bs .11956&\bs -.78175&\bs .49510&\bs -.29766 \\

Interp. (T16)&\bs 0 & \bs .63395&\bs  -.20433&\bs .11458&\bs  -.07387&
\bs  .05001&\bs  -.03428&\bs .02333 & \bs  -.01555\\

$\partial/\partial x$ (P16)&\bs 0&\bs .94118&\bs -.39216&\bs .19264&\bs -.09391
&\bs  .04293&\bs -.01789&\bs  .00667&\bs -.00219 \\

$\partial/\partial x$ (P24)&\bs 0&\bs .96000&\bs -.42462&\bs .23066&\bs-.12974
&\bs .07158 &\bs -.03778&\bs .01880 &\bs -.00874 \\


$\partial/\partial x$ (FFT)&\bs 0&\bs1.0000 &\bs-.50000&\bs.33333&\bs-.25000
&\bs .20000  &\bs-.16667 &\bs .14286&\bs-.12500\\

$\partial/\partial x$ (Pad\'e)&\bs 0&\bs .87910&\bs-.35546&\bs.13577&\bs-.05186
&\bs  .01981 &\bs -.00757&\bs  .00289&\bs-.00110\\


\hline
Operation &\bs&\bs $M_9$&\bs $M_{10}$&\bs $M_{11}$&\bs $M_{12}$&\bs $M_{13}$&
\bs $M_{14}$&$M_{15}$&$M_{16}$\\
\hline

$\partial/\partial x$ (T16)&  &\bs .16517&\bs -.80806&\bs .30987&\bs -.48958
&\bs -.65096&\bs .86017&\bs -.69157&\bs .44690\\

Interp. (T16)&  &\bs  .01005&\bs  -.00622&\bs   .00365&\bs  -.00199
&\bs  .00099&\bs  -.00043&\bs  .00015&\bs  -.00002\\

$\partial/\partial x$ (P16)& &\bs .00062&\bs -.00015&\bs .00003&\bs $\sim 0$
&\bs $\sim 0$ &\bs $\sim 0$ &\bs $\sim 0$ &\bs $\sim 0$ \\

$\partial/\partial x$ (P24)& &\bs  .00377&\bs -.00150&\bs  .00054&\bs-.00018
&\bs  .00005 &\bs -.00001 &\bs $\sim 0$ &\bs $\sim 0$ \\


$\partial/\partial x$ (FFT)& &\bs .11111 &\bs-.10000 &\bs .09091 &\bs-.08333
&\bs .07692 &\bs-.07143 &\bs .06667&\bs -.06250\\

$\partial/\partial x$ (Pad\'e)& &\bs .00042 &\bs-.00016&\bs .00006&\bs-.00002
&\bs  .00001 &\bs $\sim 0$&\bs $\sim 0$&\bs $\sim 0$\\

\hline
\end{tabular}
\end{center}

{\it Table \tablemm: Finite-difference coefficients for large stencil radii,
with the same notation as for table \tablem. The label ``FFT" denotes the
coefficients for a Fourier transform (section \ref{compact}).}

\begin{figure}[!] \plotone{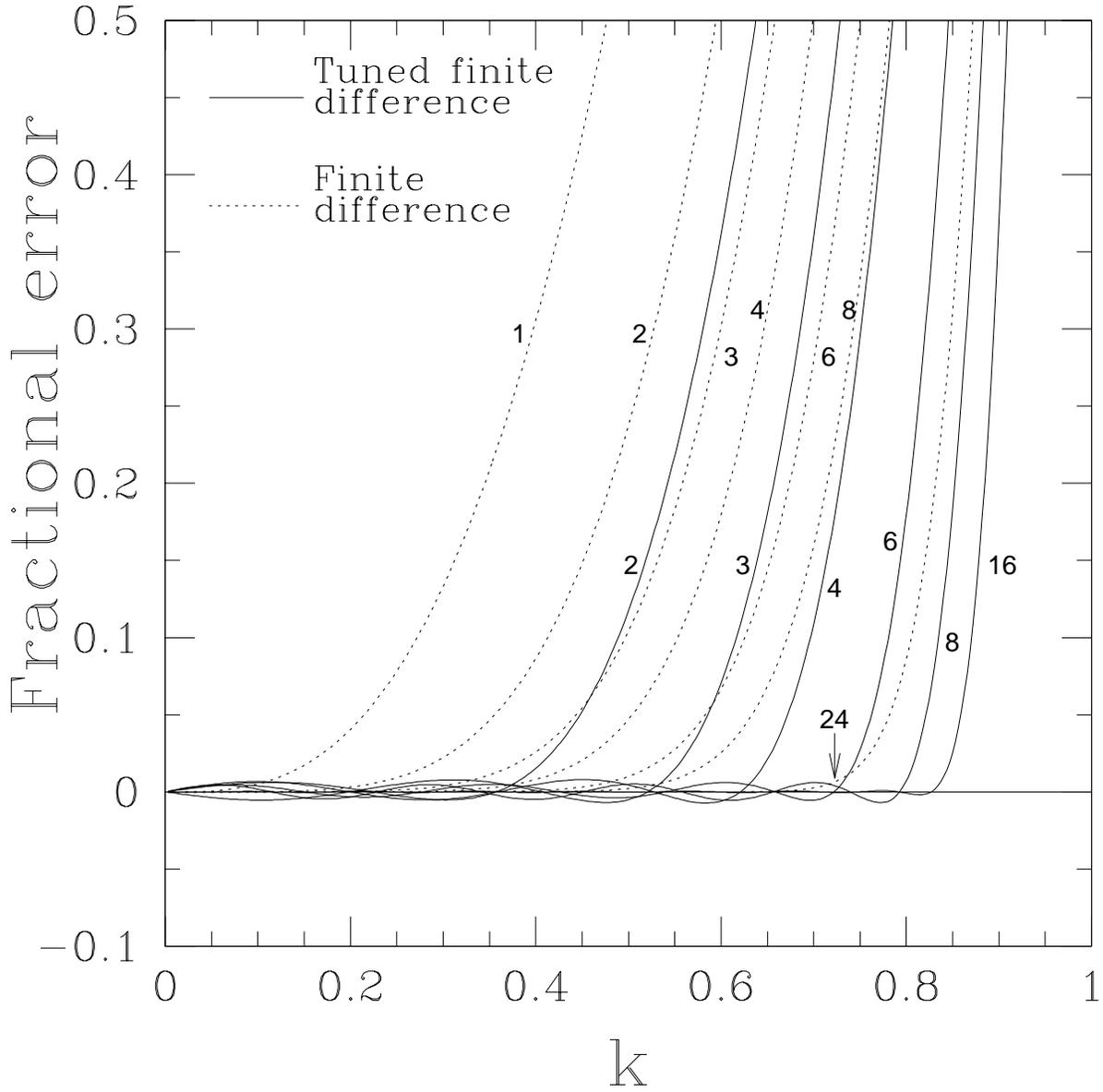} \caption{\it Figure \figerror:
The relative gradient error for the polynomial-based finite difference (dotted
line) and tuned finite difference (solid line). Numbers indicate the stencil
radius.}  \end{figure}

\subsection{Compactness} \label{compact}

The Pad\'e scheme evaluates implicit derivatives: \begin{equation} f_i^\prime +
\sum_{j=1}^{p^\prime} P_j^\prime (f_{i+j}^\prime + f_{i-j}^\prime) =
\sum_{j=1}^{p} P_j (f_{i+j} - f_{i-j}) \end{equation} The case $p^\prime=1$
involves a cyclic tridiagonal matrix, $p^\prime=2$ a cyclic pentadiagonal
matrix, etc.  The Pad\'e scheme is not compact because the gradient evaluation
draws information from all grid points, propagated around the grid by the
cyclic matrix. The contribution from each grid point can be explicitly
evaluated from the matrix inverse to express it in the form of equation
\ref{stencil}.  The 6th order tridiagonal scheme with ${P_1^\prime=1/3,
P_1=7/9, P_2=1/36}$ yields the coefficients denoted ``Pad\'e" in table
\tablemm, where the entries $M_j$ have significant magnitudes up to $j=6,$ in
spite of the fact that $P_j$ extends only to $j=2.$ The Pad\'e scheme has a
maximum wavenumber of $K=.50$ whereas the radius-6 tuned scheme has $K=.74.$
The tuned scheme is maximally compact in the sense that it has the best
wavenumber-accuracy properties if the information to be drawn from is confined
to within a given stencil radius.

The spectral gradient is particularly noncompact. The coefficients for a size
$N$ transform expressed in the form of equation \ref{stencil} are
\begin{equation} M_j = \frac{4 \pi}{N^2} \sum_{s=1}^{N/2} s \sin(2 \pi s j / N)
\,\,\,\, \longrightarrow \,\,\,\, j^{-1} \,\,\,\, (N \rightarrow \infty).
\end{equation} It is interesting to note that these are the coefficients for
the linear function expressed as an infinite sine series. Unfortunately, quite
a few terms are needed before the series starts to resemble the linear
function.

\subsection{Higher-order gradients and other operations}
\label{interpolateoned}

Diffusion involves second-order gradients. For these, we utilize the cosine
modes to construct an error function analogous to equation (\ref{eqerror}):
$$E_S^{(2)}(k) = (k \pi)^2 + M_0 + 2 \sum_{j=1}^S M_j cos(\pi k j).$$ For
fourth-order gradients, $$E_S^{(4)}(k) = - (k \pi)^4 + M_0 + 2 \sum_{j=1}^S M_j
cos(\pi k j).$$ A low pass filter can mimic a high-order hyperdiffusivity. To
construct this, replace the $(\pi k)^2$ term with something that is zero at low
$k$ and large at high $k$ with a sharp transition at the desired cutoff $k$.

An interpolation halfway between two grid points is useful for resolution
refinement and dealiasing: \begin{equation} E_S^{(I)}(k) = - \pi + 2
\sum_{j=1}^S M_j cos(\pi k (j-1/2)). \label{eqinterp} \end{equation} Here,
$j=1$ corresponds to the nearest point, $j=2$ to the next most distant point,
etc.

\section{Resolving power of a turbulence simulation} \label{flop}

We address two elements of computational efficiency - execution speed and
resolution per grid element, which together constitute the resolving power.
Other factors exist that won't be considered here, such as Lagrangian
vs. Eulerian timesteps, and variable timesteps.

Define the execution speed as the number of grid zones (or particles, for an
SPH code) processed per CPU GigaFlop per second, or Kilo-Elements per GFlop per
second (Kegs).  Also define a resolution per grid element ($K$) as a measure of
the effective resolution of a grid element or particle: $$ K = \frac{k_{Max}}
{k_{Nyquist}} $$ where $k_{Max}$ is the maximum wavenumber for which accurate
gradients can be computed and $k_{Nyquist}$ is the Nyquist wavenumber, the
maximum wavenumber expressible on the grid. $k_{Nyquist} = \pi / \Delta$ for a
grid spacing of $\Delta.$ The value of $K$ for various finite difference
schemes appears in table \tablek.

The resolving power (R) is the measure of the speed of a code at a fixed
benchmark resolution. For example, a code with high $K$ requires fewer grid
cells and hence executes faster than a code with lower $K$. The scaling for a
3D algorithm based on explicit time differences of second-or-higher order is
$$R = [\mbox{Kegs}] \cdot K^3.$$ For a flux-transport algorithm, it is $R =
[\mbox{Kegs}] \cdot K^4.$

Many algorithms such as the Riemann shock solver aren't based on explicit
gradients and lack an easily definable $K.$ $K$ can alternatively be defined in
terms of the maximum achievable Reynolds number $R_e$ on an $N^3$ grid: $R_e =
(A N K)^{4/3}.$ Maron \& Cowley (2001) found that a $128^3$ spectral simulation
with a $K = 2/3$ dealiasing truncation had $R_e = 2500$ implying $A \sim 4.1.$
For algorithms with intrinsic numerical viscosity plus an explicit Laplacian
viscosity, the value of the viscosity parameter can be varied to identify the
minimum meaningful viscosity and from that an effective Reynolds number
follows.

\subsection{Transforms} \label{transforms}

A 3D spectral code is based on real-to-complex and complex-to-real Fast Fourier
Transforms (FFTs) on an $N^3$ grid. This transform requires $7.5 \hspace{1mm}
\log_2(N)$ floating point operations per grid point.  The FFT is difficult to
optimize and executes at only some fraction of the peak floating point
speed. For the purposes of wallclock execution time, we may think of the FFT as
having effectively $7.5 log_2(N) / f_{FFT}$ operations per grid point where
$f_{FFT}$ is the efficiency factor associated with algorithm.  We further
identify separate efficiency factors for the serial and parallel aspects of the
algorithm by $F_{FFT}^s$ and $F_{FFT}^p$ respectively. The FFT poses a
challenge to optimization because the 1D stage of the transform involves highly
irregular memory access (and hence memory latency), the 3D stage requires
non-consecutive array access over the slow indices, and the parallel stage
requires communication between every processor pair.  Frigo and Johnson (1998)
met this challenge in grand style with their cross-platform application
``Fastest Fourier Transform in the West (FFTW)," which is usually the best
performer on any given architecture. On the Pittsburgh Supercomputing Center's
Alpha ES45 "Lemieux" it does especially well, with $f_{FFT}^s=0.32$ and
$f_{FFT}^p=0.38$ for a $1024^3$ transform on 512 processors. For a $512^3$ grid
on 256 processors, $f_{FFT}^s = 0.32$ and $f_{FFT}^p = 0.44.$ Other large-grid,
large-CPU transforms perform similarly (figure \figfftw).  We take these
parameters as the practical limit of FFT efficiency.

\begin{figure}[!] \plotone{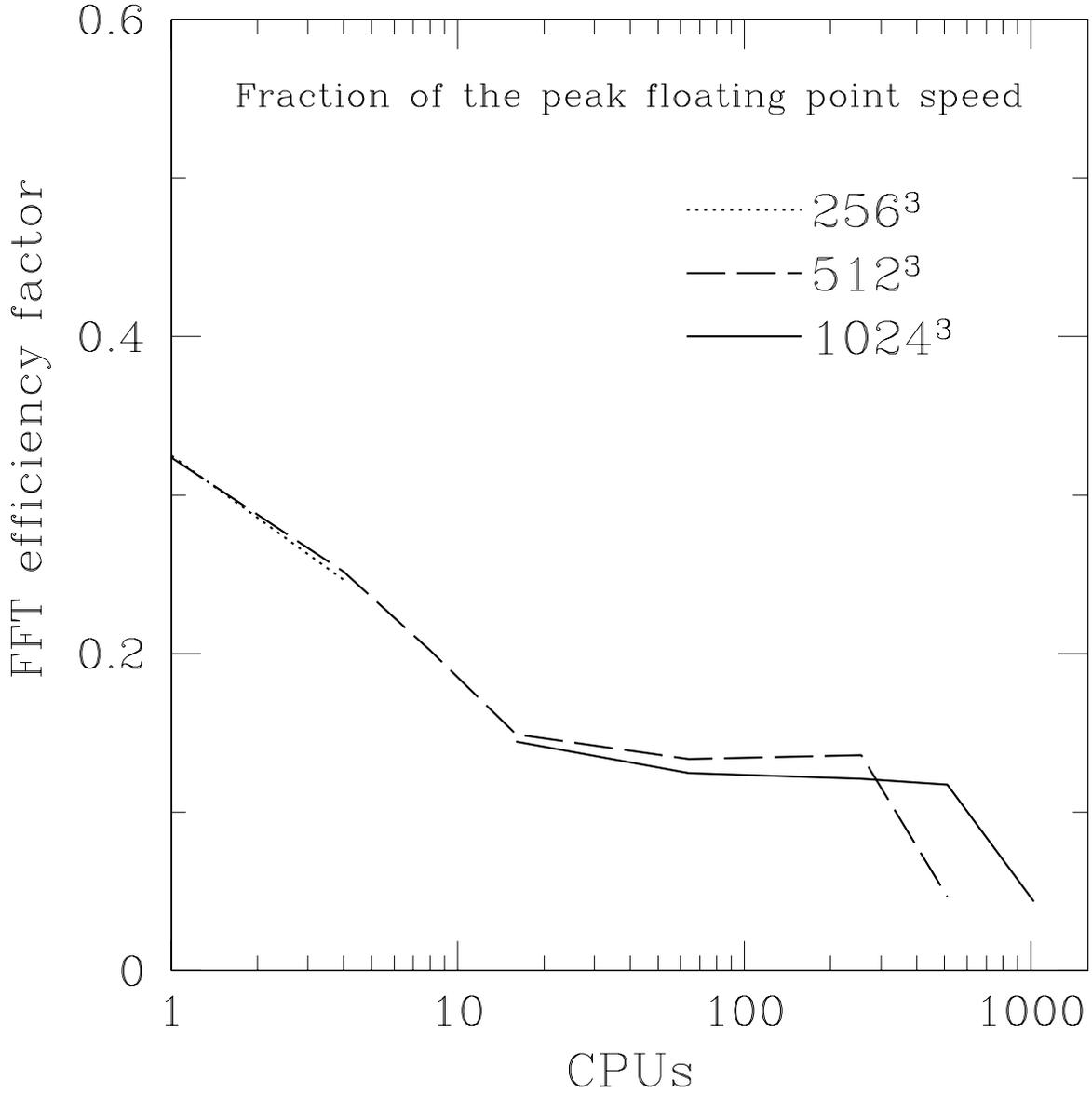} \caption{\it Figure \figfftw: The
performance of FFTW for $256^3,$ $512^3,$ and $1024^3$ transforms on the
Pittsburgh Supercomputing Center's MPI-parallel machine ``Lemieux." The
efficiency factor is the fraction of the peak floating point speed. Note that
the parallel inefficiency takes effect at a low number of CPUs and plateaus at
a high number of CPUs.}  \end{figure}

\subsection{The finite-difference gradient transform}

The finite-difference gradient transform (equation \ref{stencil}) on a radius
$S$ stencil has a floating point operation count per grid element of $(4 S -
2).$ Unlike the FFT, memory access is linear and parallel communition occurs
only between the two adjacent processors.  The transform can be optimized so
that the serial and parallel efficiency factors $f_{FD}^s$ and $f_{FD}^p$ are
much closer to unity than for the FFT case. On the Pittsburgh Supercomputing
Center's Alpha ES45 ``Lemieux", $f_{FD}^s = 0.85$ and $f_{FD}^p = 0.8$ to yield
an overall finite-difference efficiency factor of $f_{FD} = 0.68.$

\subsection{Pad\'e gradients}


The Pad\'e scheme is nonlocal - the transform has to be computed simultaneously
on a 1D cut that transverses the volume. The operation count is 5 adds and
multiplies plus one divide for the tridiagonal scheme, and 7 adds and
multiplies plus one divide for the pentadiagonal scheme (Lele 1992).  On the HP
ES45, a divide takes 12 clocks, and since it can't be pipelined the effective
cost is 24 floating point operations. This is because ordinarily the ES45
produces one add and one multiply result per clock cycle. The total number of
floating point operations is then $34/f_{Pade}$ for the tridiagonal scheme and
$38/f_{Pade}$ for the pentadiagonal scheme, where $f_{Pade}$ is the efficiency
factor of the algorithm.

The fact that the Pad\'e scheme is nonlocal eliminates the possibility of
adaptive resolution through a variable stencil size, and also degrades the
performance at boundaries.  The fact that it's a matrix operation makes it more
difficult to pipeline efficiently whereas the tuned finite-difference gradient
is a straightforward local convolution. The locality of the tuned gradient
involves communication only between adjacent processors whereas the Pad\'e
method is nonlocal and requires an all-to-all communications stage.  This
seriously degrades the Pad\'e scheme's parallel scalability, especially for
large numbers of CPUs (section \ref{parallel}).

\subsection{Operation count for one timestep}

The floating point operation count per timestep is the operation count per
transform times the number of transforms.  For the transform count, assume a
Runge-Kutta 2nd order timestep (RK2) so that the field gradients are evaluated
twice per timestep.  A spectral hydro simulation involves 3 transforms from
Fourier to real space and 6 transforms back to Fourier space, all done twice
for a total of 18 transforms. The MHD case has 30 transforms.

Aliasing error limits the spectral resolution of a bilinear partial
differential equation to $K = 2/3,$ or $K = \sqrt{8/9} = .94$ with the
inclusion of a phase shift correction (Canuto et. al. 1987).  The correction is
implemented by calculating the time derivatives on two grids - the original
grid and a grid shifted to the zone center - and averaging the results after
shifting back to the original grid. The procedure can be coordinated with the
Runge-Kutta algorithm by calculating the two Runga-Kutta stages on grids
shifted diagonally from each other by half a grid zone (Canuto et. al. 1987).
The grid shift adds negligible extra computation for the spectral algorithm
because it can be carried out in Fourier space as a mere multiply. For a
finite-difference code it costs two real-space interpolations per dynamical
field. The interpolation can be tuned for enhanced wavenumber accuracy
analogously with finite-difference gradients.  The radius-4 hypergradient is
accurate up to $K = .64$ and needs no grid-shift aliasing correction whereas
higher-order hypergradients do.

To estimate the operation count of the finite-difference method we assume the
adiabatic equations of MHD, although other forms are possible.

\begin{equation} \partial_t \V = - \V \cdot \bnabla \V - \rho^{-1} \bnabla P +
\rho^{-1} (\bnabla \times \B) \times \B + \nu \nabla^2 \V + \nu_4 \nabla^4 \V
\label{mhda} \end{equation}


\begin{equation} \partial_t \B = - \bnabla \times (\V \times \B)
+ \eta \nabla^2 \B
+ \eta_4 \nabla^4 \B
\label{mhdb}
\end{equation}



\begin{equation} \partial_t \rho = - \bnabla \cdot (\rho \V) \hspace{12mm}
P \sim \rho^\gamma \hspace{12mm}
\bnabla \cdot \B = 0
\label{mhdc} \end{equation}



\begin{center}
\begin{tabular}{|llll|}
\hline
$\V$ & Velocity field & $\B$ & Magnetic field \\
$\nu$ & Viscosity & $\eta$ & Magnetic resistivity \\
$\nu_4$&4th order hyperviscosity&$\eta_4$&4th order magnetic hyperresistivity\\
$P$ & Fluid Pressure & $\rho$ & Density \\
$\gamma$ & Adiabatic index & & \\
\hline
\end{tabular}
\end{center}

For hydrodynamics (without a magnetic field) there are 9 velocity and 3 density
gradients. A magnetic field adds 9 magnetic gradients. Diffusion terms don't
add to the computation because they only need to be applied once every few
timesteps (section \ref{diffusion}). Each gradient is evaluated twice per
timestep for Runge-Kutta order 2. The grid shifts are evaluated twice per field
per timestep for stencil radii larger than 4. These considerations determine
the number of transforms in table \tablekegs. Note that for the finite
difference and Pad\'e cases, each coordinate direction is counted as one
transform. For the FFT case, the 3D transform is counted as one transform.

\begin{center}
\begin{tabular}{|lccccccc|}
\hline
Algorithm&Physics &Stencil& Ops/   &Trans-&Speed&Max wave-&Resolving\\
         &        &radius &element &forms &(Kegs)&number& Power \\
         &        & (S)   &transform&   && (K) &$(\mbox{Kegs} \cdot K^{3})$\\
\hline
Hypergradient&Hydro& 2&   6$f_{FD}^{-1}$& 24 & 6940$f_{FD}$&.38& 381$f_{FD}$\\
Hypergradient&Hydro& 3&  10$f_{FD}^{-1}$& 24 & 4170$f_{FD}$&.54& 656$f_{FD}$\\
Hypergradient&Hydro& 4&  14$f_{FD}^{-1}$& 24 & 2980$f_{FD}$&.64& 780$f_{FD}$\\
Hypergradient&Hydro& 8&  30$f_{FD}^{-1}$& 32 & 1040$f_{FD}$&.80& 533$f_{FD}$\\
Hypergradient&Hydro&16&  62$f_{FD}^{-1}$& 32 &  504$f_{FD}$&.84& 299$f_{FD}$\\

Hypergradient& MHD & 2&   6$f_{FD}^{-1}$& 42 & 3970$f_{FD}$&.38& 218$f_{FD}$\\
Hypergradient& MHD & 3&  10$f_{FD}^{-1}$& 42 & 2380$f_{FD}$&.54& 375$f_{FD}$\\
Hypergradient& MHD & 4&  14$f_{FD}^{-1}$& 42 & 1700$f_{FD}$&.64& 446$f_{FD}$\\
Hypergradient& MHD & 8&  30$f_{FD}^{-1}$& 56 &  595$f_{FD}$&.80& 305$f_{FD}$\\
Hypergradient& MHD &16&  62$f_{FD}^{-1}$& 56 &  288$f_{FD}$&.84& 171$f_{FD}$\\
\hline
Spectral        &Hydro&N/2&             625 & 18 &   89  &.94&  74  \\
Spectral        & MHD &N/2&             625 & 30 &   53  &.94&  44  \\
\hline
Polynomial   &Hydro&1 &   2$f_{FD}^{-1}$& 24 &20833$f_{FD}$&.12&  36$f_{FD}$\\
Polynomial   &Hydro&2 &   6$f_{FD}^{-1}$& 24 & 6940$f_{FD}$&.24&  96$f_{FD}$\\
Polynomial   &Hydro&3 &  10$f_{FD}^{-1}$& 24 & 4170$f_{FD}$&.34& 164$f_{FD}$\\
Polynomial   &Hydro&4 &  14$f_{FD}^{-1}$& 24 & 2980$f_{FD}$&.40& 190$f_{FD}$\\
Polynomial   &Hydro&6 &  22$f_{FD}^{-1}$& 24 & 1890$f_{FD}$&.48& 209$f_{FD}$\\

\hline

Pad\'e tridiag.  &Hydro&N/2&34$f_{\spa}^{-1}$&24&1230$f_{\spa}$&.50&
153$f_{\spa}$\\

Pad\'e pentadiag.&Hydro&N/2&38$f_{\spa}^{-1}$&32& 822$f_{\spa}$&.83&
470$f_{\spa}$\\

Pad\'e tridiag.  &MHD&N/2&34$f_{\spa}^{-1}$& 42 &700$f_{\spa}$&
.50& 88$f_{\spa}$\\

Pad\'e pentadiag.&MHD&N/2&38$f_{\spa}^{-1}$& 56 &470$f_{\spa}$&
.83&269$f_{\spa}$\\
\hline

\end{tabular}
\end{center}

{\it Table \tablekegs: The speed, resolution, and resolving power for spectral
and finite-difference algorithms. The value of $K=.94$ for the spectral
algorithm arises from the $\sqrt{8/9}$ rule for the 3D staggered-grid
dealiasing procedure (Canuto et. al. 1987). Execution speed is estimated from
the transform time while other overhead is ignored. The finite-difference
efficiency factor is approximately $f_{FD} \sim .7$}

Assuming that the finite-difference efficiency factor $f_{FD} = f_s f_p$ can be
optimized to a better degree than for the FFT transform, the tuned gradient
method could potentially be up to 7 times faster than the spectral method.  The
most efficient hypergradient configuration is with a stencil radius of 4
because larger stencil radii require an aliasing correction. More resolution
could be achieved with a loss of efficiency by using larger-radius stencils.

\subsection{Diffusion} \label{diffusion}

Diffusion serves two roles: the removal of energy cascading to the inner scale,
and as a dealiasing filter.  If diffusion is the only agent acting in the
Navier-Stokes equation, the Fourier component $\hat{\V}$ evolves as $\partial_t
\hat{\V} = - \nu k^2 \hat{\V}.$ Defining $k_<$ to be the limiting resolution,
the inner-scale modes evolve with a timescale of $t_< \sim 1/(\nu k_<^2).$ The
diffusion $\nu$ is generally set so that $t_<$ corresponds to the dynamical
cascade timescale so that diffusion balances energy cascading to the inner
scale. This is also known as the Lagrangian timescale. The timestep, however,
is proportional to the Eulerian timescale, the time for the fluid to flow a
distance equal to the resolution scale. Generally, the Lagrangian timescale
greatly exceeds the Eulerian timescale and so diffusion has only a small effect
each timestep.  We can therefore economize by applying kinetic and magnetic
diffusion only once every few timesteps with a corresponding increase in the
values of $\nu$ and $\eta.$ This effectively removes diffusion from the
computational load. In section \ref{tests} we find that once every four
timesteps yields equivalent results as once every timestep. Furthemore, the
coefficients of the second and fourth order diffusion operators as well as
those for a dealiasing filter can be added to compactify them into one
operator.

\subsection{Magnetic divergence} \label{ct}

Magnetic divergence can be suppressed either with a scheme that preserves it
exactly, such as the constrained-transport algorithm (Stone \& Norman, 1992 I,
1992 II), or by letting it evolve unconstrained and periodically repairing it
such as with an FFT. The constrained-transport scheme derives electric field
interpolations and gradients from first first-order finite differences and
hence has poor wavenumber resolution. High-wavenumber finite differences
improve this situation and in fact magnetic divergence grows slowly enough so
as to not be concern (section \ref{tests}.)

\subsection{Parallelization} \label{parallel}

Good parallel scalability occurs if the demand for communications is less than
that for floating point arithmetic so that both can occur simultaneously,
otherwise communications hinder execution. The relevant indicator is the ratio
$R$ of floating point operations per second divided by the transmission rate of
floating point variables between processors.  For a machine such as the
Pittsburgh Supercomputing Center's 3000 processor HP Alpha ES45 ``Lemieux," the
ratio is $R_{Arch} = 2 \, \mbox{GFlops} \, / \, (68\cdot 10^6 \,
\mbox{reals/s}) \sim 30$ (San Diego Supercomputing Center study, 2003).  CPU
technology tends to improve more rapidly than communications and so this ratio
may increase over the next few years (table \tablecpu).

A finite-difference transform on a radius $S$ stencil invokes $2S-1$ adds and
$S$ multiplies, effectively totaling $4S-2$ operations since add and multiply
units appear in pairs. Summed over the three coordinate directions this is
$12S-6$ floating point operations.  For parallel execution, let the data be
distributed in slabs of dimension $(N,N,N^\prime)$ where the grid size is $N^3$
and the number of CPUs is $C = N/N^\prime.$ To compute the $z$ gradient
transform every processor summons and sends $2 N^2 S$ reals to and from
adjacent processors. The two-way MPI pass proceeds at double the one-way rate
quoted in table \ref{parallelb}.  The ratio of floating point operations to
passed variables is then $R_{FD} = N^\prime(6 - 3/S).$ Efficient parallel
scalability can occur if $R_{FD} > R_{Arch}$ or $N^\prime > 4.$

For the Pad\'e transform, the entire 1D coordinate slice has to be on the same
processor. For slab geometry, the $x$ and $y$ transforms can be done with the
data on hand and then a global interprocessor transpose reorganizes the data
along $z$ for the $z$ transform. A final transpose returns to the original slab
to calculate the time derivatives. The ratio of floating point operations to
passed variables is $R_{Pade} = 3 F / 2$ where $F$ is the number of floating
point operations per grid element in the 1D transform. With $F = 34$ for the
radius-2 tridiagonal configuration and $F=38$ for the radius-3 pentadiagonal
configuration (table \tablekegs) it's at the threshold of efficient
parallelizability.

The spectral method consists of 1D $x,$ $y$ and $z$ transforms to convert from
Fourier space to real space and back. Only one transpose per $z$ transform is
necessary because the real-space products can be conducted in the transposed
state. Therefore the ratio of floating point operations to passed variables is
$R_{FFT} = 75/f_s$ where $f_{FFT}^s \sim .32$ is the efficiency factor for the
serial stage of the computation (section \ref{transforms}). In spite of the
fact that $R_{FFT}$ is approximately 8 times $R_{Arch}$ for the supercomputer
``Lemieux", the parallel efficiency is only $f_{FFT}^p = .38,$ implying that
the all-to-all transpose is substantially slower than the peak transmission
rate would suggest. Since the Pad\'e transform has about $10$ times fewer
floating point operations than the spectral transform, it is likely that the
Pad\'e scheme would be seriously inefficient in parallel execution. Since the
finite-difference transform only involves passing between the adjacent CPUs, it
does not suffer from this problem.

\subsection{Memory} \label{memory}

A code based on tuned finite differences can calculate the entire time update
for small sections of the simulation volume at a time, eliminating the need for
large temporary storage so that only the dynamical fields $\{ \V, \B, \rho \}$
count toward the memory requirement. This is 7 arrays of size $(N,N,N^\prime)$
where $N^\prime$ is $N$ divided by the number of CPUs. It is safe to assume
that the temporary storage plus the memory taken by the operating system will
exceed the size of one array. Given that computer memory tends to come in
powers of $2,$ we assume that the total memory requirement per CPU is equal to
the size of 16 arrays of 4 Byte reals. Table \tableram \, shows how grids less
than or equal to $1024^3$ can be fit easily onto most supercomputers and a
$2048^3$ grid could fit onto a supercomputer with 512 CPUs and 1
GB/CPU. $N^\prime$ should be at least 4 for good parallel scalability (section
\ref{parallel}), and so $2048^3$ simulations are feasible on most existing
supercomputers (table \tablecpu).

A code based on Pad\'e gradients cannot be arranged to compute the time update
for subsections successively as could be done for tuned finite differences.
This is because the Pad\'e transform is nonlocal whereas the tuned finite
difference operates on a local stencil only. Hence, the entire grid has to be
updated all at once, demanding temporary storage for each partial derivative
and expanding the memory requirement beyond 16 arrays per CPU. This is also
true for a spectral code.

\subsection{Supercomputers} \label{parallelb}

\begin{center}
\begin{tabular}{
| @{\hspace{2mm}}
l @{\hspace{0mm}}
l @{\hspace{0mm}}
c @{\hspace{2mm}}
c @{\hspace{2mm}}
c @{\hspace{2mm}}
c @{\hspace{1mm}}
c @{\hspace{1mm}}
c @{\hspace{1mm}}
c @{\hspace{1mm}}
c @{\hspace{1mm}}
c @{\hspace{1mm}}
c @{\hspace{0mm}}
|}
\hline
Machine&Name& Clock&GFlops&RAM &CPUs&MPI   & Pipe   &  L1  &  L2&L3&RAM\\
       &    &(GHz) &      &(GB)&    &speed & stages &(kB)&(MB)&(MB)&speed\\
       &    &      &      &    &    &(GB/s)&        & &    &    &   (GB/s)\\
\hline
NCSA Xeon          &Tungsten&3.1& 6.2&1.5&2560&     .5   &  &  &0.5&   &   \\
ORNL Cray X1       &Phoenix &0.8&12.8&4.0& 512&    5     &  &16&2.0&   &200\\
PSC HP ES45        &Lemieux &1.0& 2.0&1.0&3016&     .25  & 4&64&8.0&   &5.2\\
ORNL IBM Power4    &Cheetah &1.3& 5.2&1.0& 864&     .25  & 6&32&1.5& 32&200\\
UIUC Intel Itanium &Titan   &0.8& 3.2&2.0& 256&      .1  &  &  &   &   &\\
UIUC Intel Pentium &Platinum&1.0& 1.0&1.5& 968&      .01 & 5&  &   &   &\\
Japan NEC   &Earth Sim.     &0.5& 8.0&2.0&5120&          &  &  &   &   &\\
Iowa Athlon        &Zephyr  &1.5& 1.0&1.0&  32&      .25 & 4&  &   &   &\\
Athlon Opteron     &        &1.6& 3.2&1.0&    &          &  &64& 1 &   &5.3\\
SGI Altrix         &        &1.3& 2.6&   &    &          &  &32&.25&  3&6.4\\

\hline
\end{tabular}
\end{center}

{\it Table \tablecpu: Academic supercomputers. For Cheetah, some CPUs have 4
GB/CPU but most have 1. The French Alpha ES45 ``Ixia" and ``Nympea" have the
same characteristics as Lemieux. Most of these numbers are from Dunigan
2003. The two-way MPI passing speed is double the tabulated one-way speed. Some
entries are missing because this kind of information is often hard to come by
on the web.}


            l1    l2   l3



\begin{center}
\begin{tabular}{|ccc|}
\hline
Grid & CPUs & Memory/CPU \\
     &      & (GB) \\
\hline
$ 512^3$ &   64 &  1/8 \\
$ 512^3$ &  128 & 1/16 \\
$1024^3$ &  128 &  1/2 \\
$1024^3$ &  256 &  1/4 \\
$2048^3$ &  128 &    4 \\
$2048^3$ &  256 &    2 \\
$2048^3$ &  512 &    1 \\
\hline
\end{tabular}
\end{center}

{\it Table \tableram: Parallel configurations and memory requirements for a
finite-difference code.}

\section{Imitating an incompressible spectral MHD code}
\label{mhd} \label{tests}

It is possible to mimic the function of an incompressible spectral code with
high-wavenumber finite differences, and with significantly more resolving power
(table \tablekegs). The principal advantage of simulating incompressible
turbulence is that the sound speed does not impact the Courant timestep
condition.  We start with the equations of MHD (equations \ref{mhda} through
\ref{mhdc}) but with $\rho$ set to unity and the pressure term deleted.  In its
place we add terms of the form $\nu_D \bnabla (\bnabla \cdot \V)$ and $\eta_D
\bnabla (\bnabla \cdot \B)$ to the fluid and magnetic equations respectively to
suppress the growth of divergence in the velocity and magnetic fields. $\nu_D$
and $\eta_D$ are artificial divergence diffusivities that will be discussed in
section \ref{divergence}.

\subsection{Divergence correction} \label{divergence}

An artificial term of the form $\partial_t \V = \nu_D \bnabla \bnabla \cdot \V$
in the Navier Stokes equation removes divergence while preserving the
solenoidal component.  Define $\hat{\V}_\parallel = \kk \cdot \hat{\V} / k$ and
$\hat{\V}_\perp = \hat{\V} - \hat{\V}_\parallel,$ and then the artificial term
has the effect of $\partial_t \hat{\V}_\parallel = - \nu_D k^2
\hat{\V}_\parallel$ and $\partial_t \hat{\V}_\perp = 0.$ It's not surprising
that the procedure is most effective for large $k$ modes. Fully removing the
divergence for all modes requires a global operation such as can be
accomplished with an FFT. This procedure is instead based on locally calculated
gradients. $\nu_D$ is set so that the divergence in the largest $k$ modes is
completely eliminated in one timestep: $$ \nu_D = \frac{1}{k_{Nyquist} \Delta
t} = \frac{1}{\pi^2 N^2 \Delta t},$$ where $N$ is the number of grid points in
each dimension and $k_{Nyquist}$ is the Nyquist wavenumber. A larger value
overcompensates for the largest $k$ modes.  This procedure serves to suppress
the growth of divergence for several timesteps, but not
indefinitely. Ultimately a complete correction involving FFTs must be applied,
especially for low $k$ modes. Figure \figdiv \, shows that divergence can be
held to less than 1 percent of the maximum by applying the FFT divergence
corrector once every 8 timesteps. At this frequency, the FFT contributes
negligibly to the computational load.

Define an energy spectrum: $<V^2>/2 = \int E_V(k) dk,$ where $E_V(k) = 2\pi k^2
<|\hat{V}(k)|^2>.$ Also define a divergence spectrum $D_V(k) = 2\pi k^2 <|\kk
\cdot\hat{\V}(k)|^2>$ and a divergence normalization spectrum
$\overline{D}_V(k) = 2\pi k^2 <k^2|\hat{V}(k)|^2>.$ The divergence
normalization serves to define a fractional divergence
$D_V(k)/\overline{D}_V(k).$ Analogously define a magnetic spectrum $E_B(k),$ a
magnetic divergence spectrum $D_B(k),$ and a magnetic divergence normalization
spectrum $\overline{D}_B(k).$

\begin{figure}[!] \plottwo{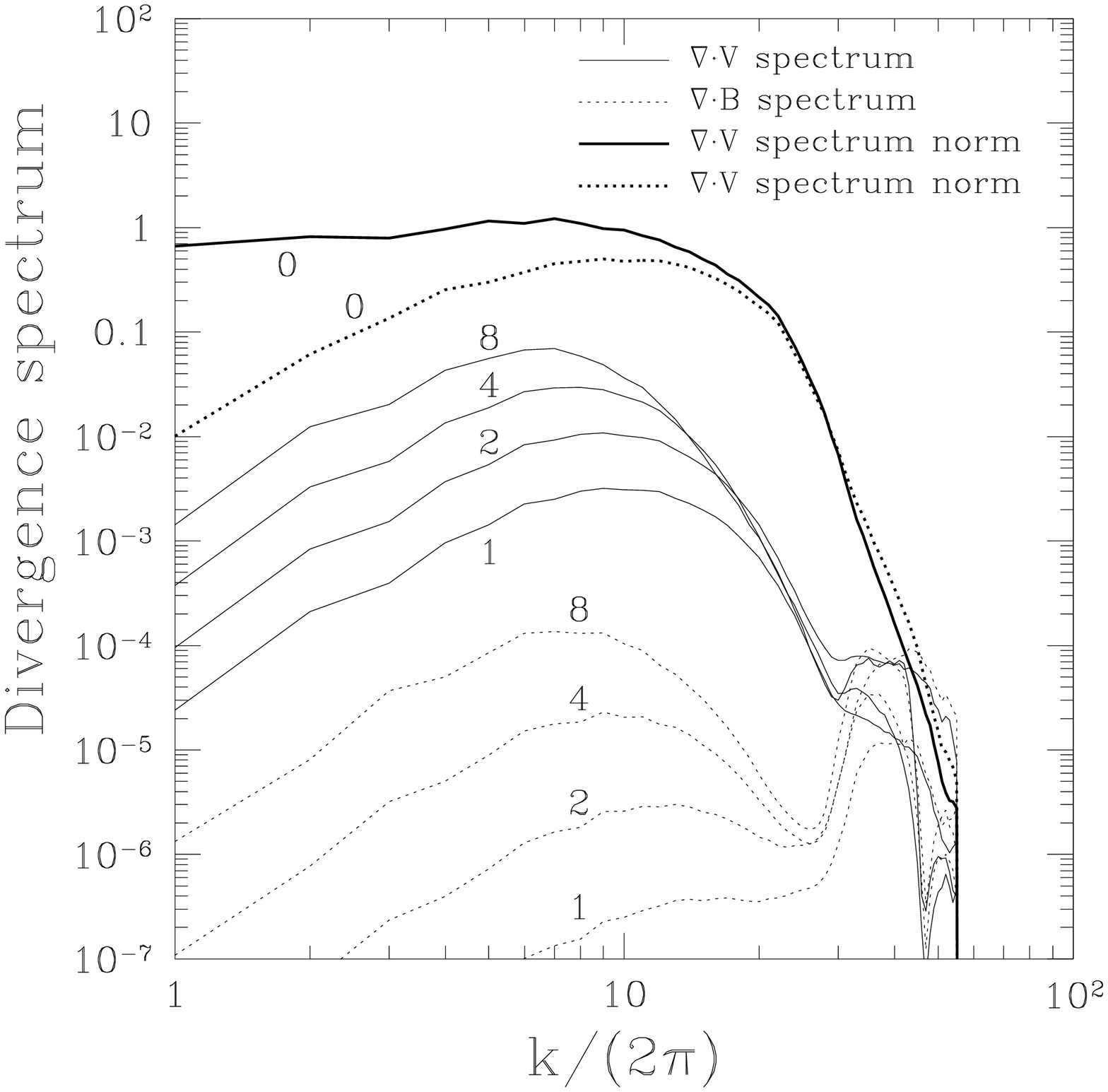}{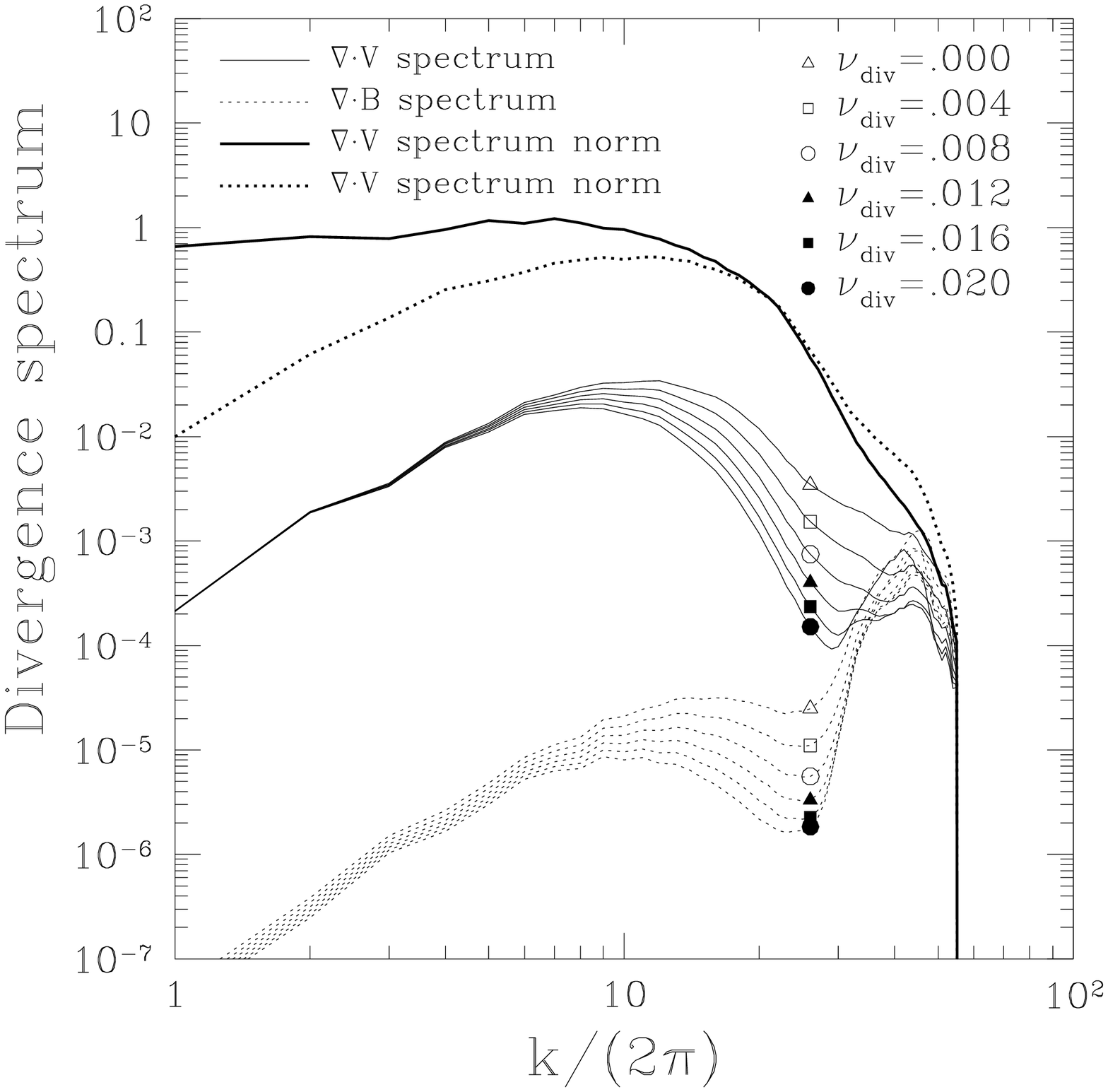} \caption{\it Figure
\figdiv: Left: The $V$ and $B$ divergence spectra for simulation Z753. Numbers
indicate timesteps. Right: The divergence spectra for simulations Z749 through
Z754 after four timesteps of evolution, for a sequence of parameters $\nu_D$
and $\eta_D.$} \end{figure}

\subsection{Diffusivity}

Neither the diffusivity operator nor the spectral divergence correction need to
occur every timestep (section \ref{diffusion}). Nearly equivalent evolution
results from applying them only once every few timesteps, lessening their
impact on the compuational load. Let the diffusivity be applied once every
$I_{diff}$ timesteps with $\nu$ and $\eta$ replaced by $\nu I_{diff}$ and $\eta
I_{diff},$ with similar replacements for the higher order diffusivities. Also,
let the spectral divergence correction be applied once every $I_{FFT}$
timesteps.

\subsection{Simulations} \label{simulations}

With the need to study both kinetic and magnetic divergence, we ran tests with
fully turbulent kinetic and magnetic fields. Initial conditions come from the
magnetic dynamo simulations of Maron \& Cowley (2001). There, forced isotropic
turbulence with a magnetic field is evolved to its long-term saturated state
where the magnetic energy equals the kinetic energy. This is the magnetic
analogue to the Kolmogorov cascade. We restarted this simulation without
forcing to study the divergence growth rate.  Table \tablesim\, lists the
simulations that were run for this study.  Z677 is a high-resolution spectral
simulation that serves as a benchmark for comparison. This simulation was run
with the spectral code ``Tulku" (Maron \& Goldreich 2001).

Simulations Z749 through Z754 (table \tablesim) were run to optimize the values
of the divergence diffusion paramters $\nu_D$ and $\eta_D.$ No spectral
divergence correction was applied ($I_{FFT} = \infty$), leaving the divergence
suppression solely to $\nu_D$ and $\eta_D.$ The initial divergence spectrum is
identically zero. The divergence spectrum after 4 timesteps is plotted in
figure \figdiv. Increasing $\nu_D$ and $\eta_D$ decreases the divergence
spectrum until $\nu_D=\eta_D=.016,$ whereupon further increase results in an
overshoot in the correction at large $k.$ Taking the value of $0.016$ as
optimal, we plot the growth of the divergence spectrum with this value
(simulation Z753) in figure $\figdiv.$ After 4 timesteps, the ratio of the
divergence to the divergence normalization spectrum is at most $0.015$ for $\V$
and $10^{-5}$ for $\B.$

In figure \figspectra\, we plot the evolution of the finite difference
simulation Z748 together with the benchmark spectral simulation Z677, both
starting from the same initial conditions. After 0.5 dynamical times, the
spectra (figure \figspectra) and real space fields (figure \figv) are in good
agreement. The spectrum for simulation Z748 is slightly diminished from Z677
because the divergence takes with it a small measure of energy. Simulation Z748
has $I_\nu = I_\eta = 4$ and $I_{FFT} = 4.$ Another simulation (Z757) has
$I_\nu = I_\eta = 1$ and $I_{FFT} = 4$ and has exactly the same energy spectrum
as Z748, establishing that the diffusion operator doesn't have to be applied
every timestep.

\begin{center}
\begin{tabular}{|cccccccc|}
\hline
Simulation& Grid  & Algorithm& $\nu_D$&$\eta_D$  &$I_\nu$&$I_\eta$&$I_{FFT}$\\
\hline
Z677  & Spectral     &$128^3$&   0    &    0     &  1       &   1   &   1\\
Z745  & Hypergradient&$ 64^3$& .016   &   .016   &  1       &   1   &   1\\
Z746  & Hypergradient&$ 64^3$& .016   &   .016   &  2       &   2   &   2\\
Z747  & Hypergradient&$ 64^3$& .016   &   .016   &  3       &   3   &   3\\
Z748  & Hypergradient&$ 64^3$& .016   &   .016   &  4       &   4   &   4\\
Z749  & Hypergradient&$ 64^3$& .000   &   .000   &  4       &   4   &$\infty$\\
Z750  & Hypergradient&$ 64^3$& .004   &   .004   &  4       &   4   &$\infty$\\
Z751  & Hypergradient&$ 64^3$& .008   &   .008   &  4       &   4   &$\infty$\\
Z752  & Hypergradient&$ 64^3$& .012   &   .012   &  4       &   4   &$\infty$\\
Z753  & Hypergradient&$ 64^3$& .016   &   .016   &  4       &   4   &$\infty$\\
Z754  & Hypergradient&$ 64^3$& .020   &   .020   &  4       &   4   &$\infty$\\
Z756  & Hypergradient&$ 64^3$& .016   &   .016   &  4       &   4   &   1    \\
Z757  & Hypergradient&$ 64^3$& .016   &   .016   &  1       &   1   &   4    \\
\hline
\end{tabular}
\end{center}

{\it Table \tablesim: Index of simulations. All simulations have
$\nu=\eta=10^{-3}$, $\nu_4=\eta_4=2.5\cdot10^{-8},$ and $\Delta t = 0.003,$
except for Z677, which has $\Delta t = 0.0012.$ $I_{FFT} = \infty$ indicates
that the FFT divergence correction is never applied.  All finite-difference
simulations utilize a radius-8 stencil and a phase-shift dealiasing
correction.}

\begin{figure}[!] \plottwo{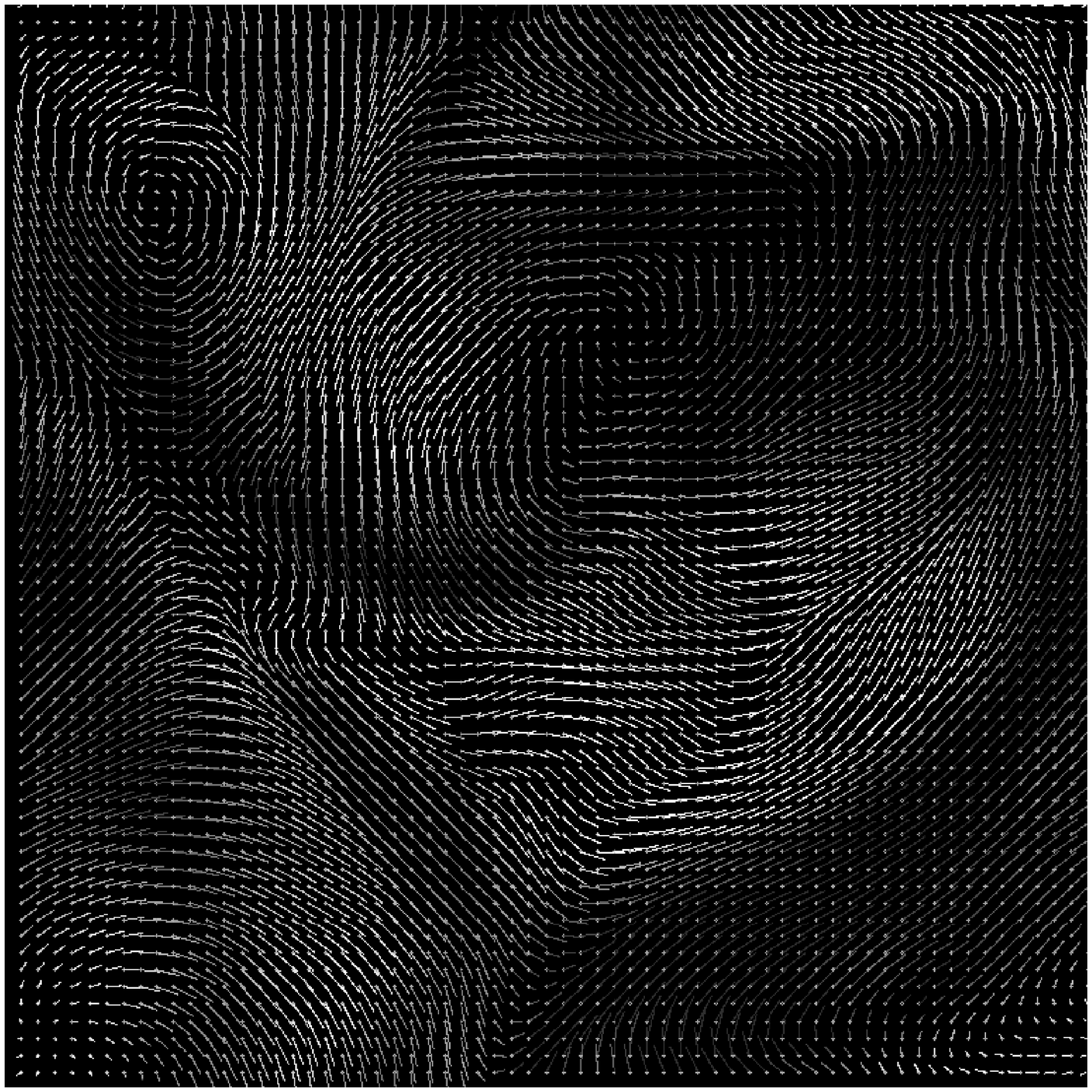}{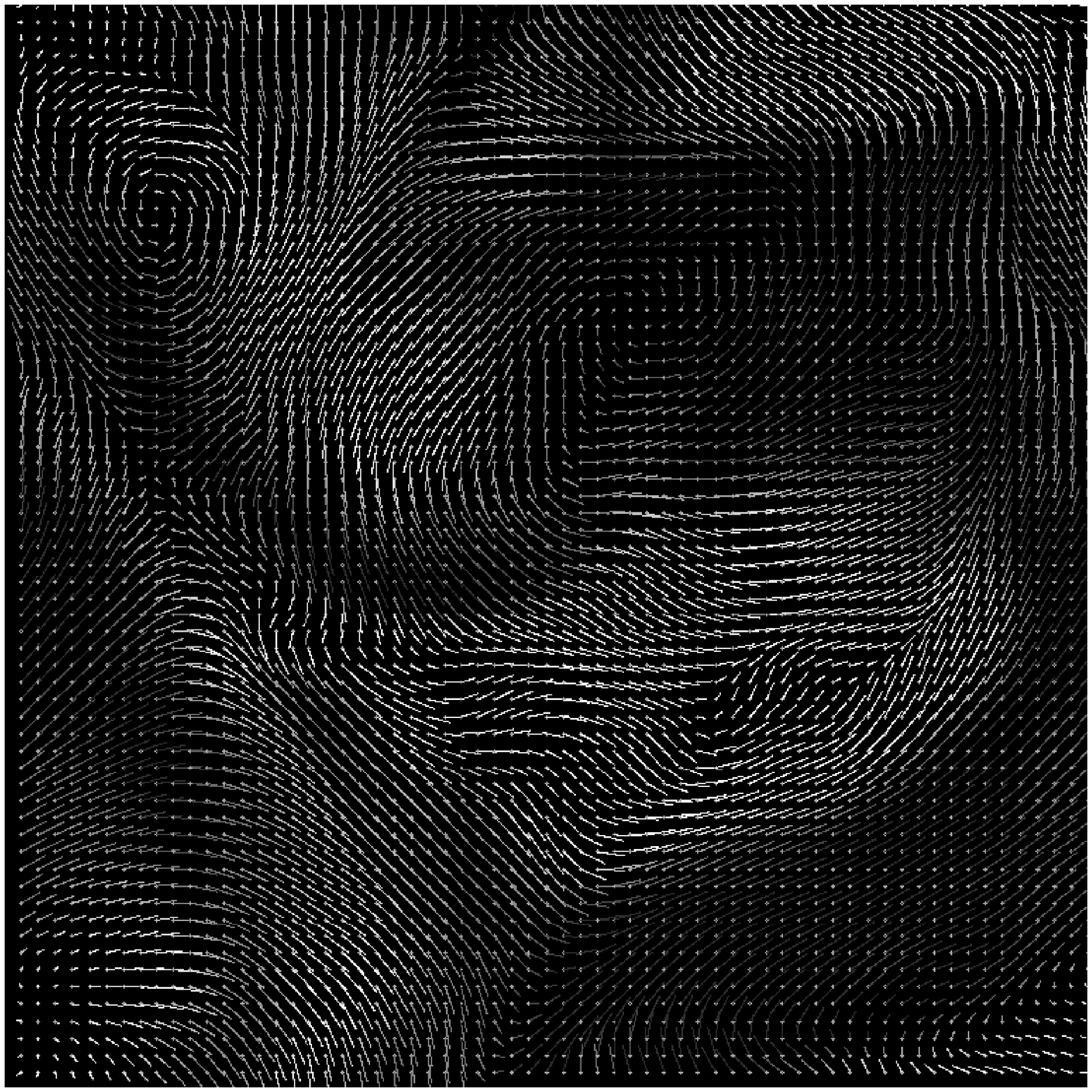} \caption{\it Figure
\figv: The velocity field in a plane slice after t=.5 for the tuned gradient
simulation Z748 (left) and the spectral simulation Z677 (right). The arrow
length gives the in-plane velocity and the arrow brightness indicates the
component transverse to the plane.}  \end{figure}

\begin{figure}[!] \plotone{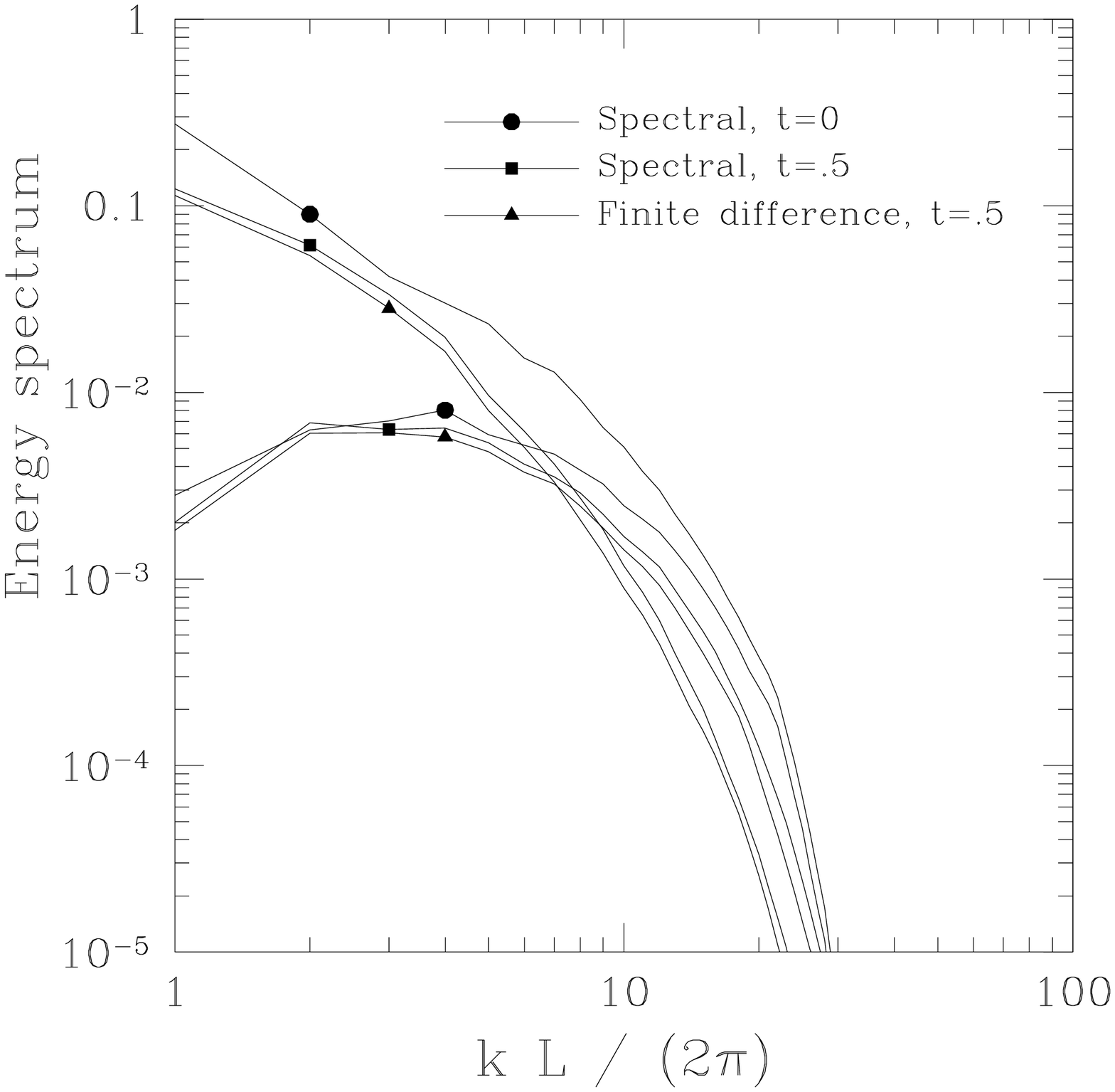} \caption{\it Figure \figspectra: The
spectra for simulations Z748 and Z677, described in section \ref{simulations}.
$k$ is the wavenumber and $L$ is the cube side length.}  \end{figure}

\section{Applications: Interpolation, refinement, and data analysis}
\label{applications}

The most accurate and most expensive interpolation procedure is drect
evaluation of the Fourier series.  Tuned finite differences provide a less
expensive interpolation high-wavenumber interpolation. For example, in 2D, the
centered interpolation (equation \ref{eqinterp}) provides function values
halfway between grid points, and another interpolation along the diagonal
yields the values at the grid centers.  We have thus doubled the resolution of
the grid, which we can do again if we wish. Note that we can do this for the
entire grid or just for a subsection. After a few doublings, a simple linear
interpolation serves to provide the function value at any point in space,
yielding a 2D interpolation with the same wavenumber resolution as the
component 1D interpolation.  This procedure generalizes to arbitrary dimension.

As if it wasn't enough trouble to run large simulations on thousands of cpus,
one is next confronted with analyzing large data cubes that are too big to fit
in the memory of one machine.  Tuned operators allow for the construction of
local local alternatives to global functions like the FFT.  These include
derivatives, dealiasing, filtering, and grid interpolation.  Large output files
can be stored in small easy-to-manage segments and operated on successively.
For the purpose of data analysis, we have provided radius-16 tuned operators in
table \tablemm\, that are accurate to 0.3 percent.

\section{Acknowledgements}

We thank Eric Blackman, Benjamin Chandran, Peggy Varniere, Yoram Lithwick, and
Tim Dennis for useful discussions, and Yoram Lithwick for the FFT benchmark
results. The simulations were run at the Pittsburgh Supercomputing Center's
Alpha ES45 ``Lemieux" with the National Resource Allocation Committee grant
MCA03S010P, and at the National Center for Supercomputing Applications' Intel
Itanium ``Titan" with the grant AST020012N. The author was supported by
Dr. Eric Blackman's DOE grant DE-FG02-00ER54600 at the University of Rochester,
and by Dr. Benjamin Chandran's NSF grant AST-0098086 and DOE grants
DE-FG02-01ER54658 and DE-FC02-01ER54651, at the University of Iowa.

\end{document}